%% file: main.tex
\documentclass[10pt,letterpaper]{article}

\usepackage{cogsci}
\usepackage{pslatex}
\usepackage{apacite}
\usepackage{caption}
\usepackage{subcaption}
\usepackage{graphicx}
\usepackage{libertinust1math}
\usepackage{bbm}

\title{Homophily and Incentive Effects in Use of Algorithms}

\cogscifinalcopy % Uncomment this line for the final submission 

\author{{\large \bf Riccardo Fogliato (rfogliat@andrew.cmu.edu)} \\
  Department of Statistics and Data Science, Carnegie Mellon University
  \AND {\large \bf Sina Fazelpour (s.fazel-pour@northeastern.edu)} \\
 Department of Philosophy and Religion \& Khoury College of Computer Sciences, Northeastern University 
  \AND {\large \bf Shantanu Gupta (shantang@andrew.cmu.edu), Zachary Lipton (zlipton@cmu.edu)} \\
  Machine Learning Department, Carnegie Mellon University
    \AND {\large \bf David Danks (ddanks@ucsd.edu)} \\
  Halicio\u{g}lu Data Science Institute and Department of Philosophy, University of California, San Diego}

\begin{document}

\maketitle

\begin{abstract}
\input{abstract}
\end{abstract}

\section{Introduction}\label{sec:intro}
\input{sections/introduction}
\section{Related Work}\label{sec:background}
\input{sections/background}

\section{Experiment}\label{sec:experiment}
\input{sections/design}

\subsection{Results}\label{sec:results}
\input{sections/results}

\section{General Discussion}\label{sec:discussion}
\input{sections/discussion}

%\section{General Discussion}

\bibliographystyle{apacite}

\setlength{\bibleftmargin}{.125in}
\setlength{\bibindent}{-\bibleftmargin}

%\pagebreak
\bibliography{refs}

\end{document}

%% file: abstract.tex
As algorithmic tools increasingly aid experts
in making consequential decisions,
the need to understand the precise factors 
that mediate their influence 
has grown commensurately. 
In this paper, we present a crowdsourcing vignette study
designed to assess the impacts of two plausible factors 
on AI-informed decision-making. 
First, we examine homophily---\emph{do people 
defer more to models that tend to agree with them?}---by 
manipulating the  agreement during training 
between participants and the algorithmic tool. 
Second, we considered  incentives---\emph{how
do people incorporate a (known) cost structure 
in the hybrid decision-making setting?}---by 
varying rewards associated with true positives vs. true negatives. 
Surprisingly, we found limited influence 
of either homophily and no evidence 
% of incentives, 
of incentive effects,
despite participants performing similarly to previous studies. 
Higher levels of agreement between the participant and the AI tool 
yielded more confident predictions, 
but only when outcome feedback was absent.
These results highlight the complexity 
of characterizing human-algorithm interactions, 
and suggest 
% Z: weird.. we don't exactly "transfer" findings
% limits on transferring findings 
% from social psychology into settings with algorithms. 
that findings from social psychology 
may require re-examination when
humans interact with algorithms.

%% file: sections/introduction.tex
In a variety of sensitive domains,
predictive models and algorithms have been adopted 
to aid expert decision makers.
This trend has motivated an extensive body of work
aimed at characterizing the performance of these models
with respect to their accuracy, fairness, 
and a variety of other desiderata. 
However, the majority of this work 
has focused on the models in isolation.
More recently, researchers have recognized
that the quality of AI-assisted decisions 
critically depends on human cognition: 
How do the relevant humans interpret, 
process, and integrate these predictions
into their decision-making processes?
% TODO: add citations

Prior work on human use of algorithmic recommendations 
has largely focused on the decision context~\cite{kleinberg2018human,dietvorst2015algorithm,de2020}, 
or features of decision subjects~\cite{green2019principles}. 
In this paper, we instead focus on the ``social'' interactions 
between humans and algorithms: 
Do people regard algorithms similarly to other people?

Social psychological research 
(e.g.,~\citeA{hoppitt2013social,turner1991social})
has revealed two factors that frequently 
shape social learning and interactions. 
First, people use a variety of heuristic strategies 
to estimate the reliability of an information source, 
and hence the value of the information it provides~\cite{hoppitt2013social,kendal2018social}. 
One common heuristic is based on \textit{homophily}: 
People use the \emph{perceived similarity}
of an information source to themselves  
as an indicator of the source's reliability. 
In social interactions, 
these perceptions of similarity 
are moderated by factors such as agreements 
in prior interactions \cite{o2019misinformation, zollman2015modeling}
as well as deeper identity-based considerations, 
e.g., shared disciplinary backgrounds \cite{phillips2006surface} 
and sociocultural identities \cite{turner1991social}. 

Second, people's decision \emph{uncertainty} 
\cite{toelch2015informational,kendal2018social}
and \emph{costs} \cite{hoppitt2013social,boyd1988culture} 
are thought to moderate reliance 
on \textit{social} (as opposed to individual) learning.
Information from social sources can be particularly critical
when navigating novel, ambiguous, risky, or high-consequence environments. 
More generally, there is evidence that incentive structures 
may shape people's predictive and decision-making behavior~\cite{evans2004if}.

While people's use of information from other humans 
is shaped by homophily and perceived costs, 
it is unknown whether people conceptualize algorithms 
in ways that would produce similar influences 
(though see \citeA{lu2021human}). 
Both of these factors can be characterized in information-theoretic terms, 
so they plausibly might extend to algorithms as well. 
We thus conducted an experiment that explicitly manipulated these factors 
in the human-AI interaction context,
including measures of decisions and perceptions. 

%% file: sections/background.tex
Our experiment engages with several bodies of prior research.
First, many researchers have aimed to characterize the various
sources and types of biases
that arise throughout the machine learning 
% learning 
pipeline
\cite{fazelpouralgorithmic, mitchell2021algorithmic}, 
and in the uptake of algorithmic information 
by users in particular 
\cite{logg2019algorithm,dietvorst2015algorithm,mosier1998automation}. 
Previous research has identified factors 
that can result in biases of overreliance
(automation bias; e.g., \citeA{skitka2000accountability}) 
or instead underreliance 
(algorithm aversion; e.g., \citeA{dietvorst2015algorithm})
on algorithmic recommendations. 
For example, the perceived difficulty of understanding 
the algorithm \cite{yeomans2019making}
or use of sensitive information about decision subjects, 
such as race \cite{green2019disparate}
or socioeconomic status \cite{skeem2020impact}, 
can influence use of algorithmic predictions.

Second, we connect with research on the proper design of algorithmic tools. 
For example, \citeA{duan2020does} are inspired by work 
that highlights the benefits of team diversity 
to find ways to counteract biases in crowdwork. 
The advantages of complementarity in teams has also
motivated work on the design of decision support tools 
that best complement human capabilities 
\cite{wilder2020learning, kamar2016directions, bansal2021most}.
Similarly, research on determinants of trust in organizational settings 
play an increasingly key role in understanding factors 
that shape human trust in algorithmic tools \cite{glikson2020human}. 
Prior work by \citeA{lu2021human} 
is the closest in approach to our study. 
\citeA{lu2021human} also examine and find evidence 
for the use of \textit{agreement} 
as a proxy for algorithmic reliability by human users,
though they focus on human-AI disagreement on cases 
where humans are confident in their predictions.
In addition, they provide only binary algorithmic outputs.
We instead focus on low-confidence cases 
and provide the model's likelihood estimates, 
which help human trust calibration \cite{zhang2020effect}.

% for Sina: introduce 
Third, our study belongs 
to a line of work that aims to extrapolate 
general human behavioral patterns
in the presence of algorithmic tools 
via crowdsourcing experiments \cite{fogliato2021impact}.
%While questions remain about the generalizability 
%of the findings from these crowdsourcing studies, 
%especially to high-stakes domains, 
%these experiments have provided some initial guidance. 
Much of this work has focused on understanding 
how human trust and reliance  
can vary with the underlying properties of these tools,
and also with the type of algorithmic recommendations 
that are communicated 
\cite{yin2019understanding, zhang2020effect}, 
such as the presence of explanations 
\cite{bansal2021does, dodge2019explaining, poursabzi2021manipulating, lai2019human, wang2021explanations}. 
Other common themes include the impact of the tools 
on the predictive and fairness properties
of human predictions \cite{green2019principles, green2019disparate}. %Our
%design choices draw from the findings of prior work across many dimensions. 
% We recall two of the main findings from prior work that directly informed the
%design choices of our experiment. 
% First, \citeA{zhang2020effect} found that study
% participants better calibrated the trust in the tool when its confidence
% scores were provided. 
Our two-step elicitation process is inspired 
by the findings of \citeA{buccinca2021trust},
who concluded that eliciting predictions from participants 
before revealing the model's recommendation 
decreased their overreliance on the tool. 
Similarly, \citeA{green2019principles} noted
that  participants achieved higher predictive performance 
when they were asked to pre-register 
their predictions made without the model. 
We draw on these lessons in the design of our experiment.

%% file: sections/design.tex
The experiment was designed to test  three key hypotheses: 
\begin{itemize}
    \item \textbf{[H1]} \textit{Influence of agreement on trust and reliance.}
    Higher agreement between algorithmic recommendations and the participant,
    particularly on cases with high predictive uncertainty, 
    will increase the participant's
    subsequent trust in, and reliance on, the model.
    \item \textbf{[H2]} \textit{Influence of incentives.}
    The incentive structure will have influences on both: 
    (a) participants' predictions, which will be skewed 
    towards outcomes with higher monetary incentives; 
    and (b) participant's reliance on the algorithm, 
    which will increase as overall costs of error increase. 
    \item \textbf{[H3]} \textit{Influence of feedback on agreement-driven
    reliance.} Receiving outcome feedback will reduce the impact of agreement as a
    reliability approximation heuristic, and so will reduce participant's reliance on the model.
    % TODO: check that there is no difference in the observed model's accuracy
    % across similarity conditions
\end{itemize}
We make a further prediction that is not central to the study design 
(so is investigated through exploratory analyses): 
%\begin{itemize}
    % \item Algorithm-participant agreement provides a heuristic for approximating model reliability, so participants may rely more on similar models when error costs are higher. 
    %\item 
Higher level of agreement may lead participants 
to be more confident about their predictions, 
especially when outcome feedback is absent. 
%\end{itemize}
We tested these hypotheses through a vignette study 
in which participants interacted with recommendations 
generated by two different algorithmic tools. 

% describe the hypotheses -- fix from here
\subsection{Method}

\subsubsection{Algorithm and vignette construction}
%% data processing
The vignettes for the experiment all involved 
descriptions of criminal defendants. 
Participants and algorithms aimed 
to predict their future re-arrest outcomes. 
The vignettes were populated using a dataset of defendants 
sentenced in Pennsylvania's federal criminal courts
between 2004 and 2006 ($N=117,464$),
including whether they were rearrested 
in the three years following release from prison 
or imposition of community supervision  \cite{fogliato2021impact}.\footnote{Criminal justice data in the U.S. are highly biased, and heavily affected by measurement issues \cite{bao2021s,
fogliato2021validity, pierson2020large, goel2016precinct}. 
In particular, re-arrest and re-offense are only imperfectly correlated, 
and those biases are (unavoidably) reflected in the AI tools trained and tested on such data
\cite{fogliato2020fairness}. 
In an effort to minimize these issues, 
we had both participants and algorithms predict
only the directly observable (though highly biased) outcome of re-arrest.} 
We focused on a subset of $N=3,523$ defendants 
for possible inclusion in a vignette,
stratified for race, sex, age, and re-arrest 
(and assuming reasonable numbers in each group).

%% AI model TODO: fix validation vs test set naming conventions
In experimentally %examining the effects of homophily via 
manipulating the levels of human-AI agreement, %the experiment required cases where either the two models agreed or they both disagreed. Moreover,
we needed to ensure that disagreements 
are not perceived as an indicator of inaccuracy 
(e.g., if an algorithm disagrees with a user over trivial cases). 
We were thus particularly interested in using cases 
of human-AI agreement (disagreement) 
that typically produced high (low) 
confidence judgments from humans. 
We identified the cases in the dataset of \citeA{fogliato2021impact} 
where more than 80\% (less than 60\%) of the participants 
made the same prediction (\textit{high (low) confidence}).

Moreover, to ensure that we used realistic algorithmic predictions, we developed two different predictive models that had comparable overall performance while occasionally disagreeing on a subset of cases (that could be used to manipulate agreement). We used stratified $70/30$ train/test sets of the full dataset (minus our presentation subsample), and trained a model using 
XGBoost \cite{chen2016xgboost}, and another using logistic Lasso
\cite{tibshirani1996regression}. Both models take defendant's demographics, the current charge type, and criminal history information as input. The probability values outputted by the models were converted into binary predictions
using a threshold of $0.5$. On the test set, the two models were well-calibrated
and had virtually identical predictive performance (AUC of $0.71$ and accuracy of $66\%$). This performance is comparable, if not superior, to the performance of many models
deployed in real-world criminal justice settings 
\cite{desmarais2020predictive}. 

% The experiment requires manipulating the extent of agreement between human users and algorithms on ``hard'' cases (where social learning is supposed to be at play). To do so with realistic algorithmic predictions, we developed two different predictive models that had comparable overall performance while occasionally disagreeing on ``hard'' cases. We used stratified $70/30$ train/test sets of the full dataset (minus our presentation subsample), and trained a model using 
%gradient boosted trees (XGBoost) \cite{chen2016xgboost}, and another using logistic Lasso
%\cite{tibshirani1996regression}. Both models take defendant's demographics, the current charge type, and criminal history information as input. The probability values outputted by the models were converted into binary predictions
%using a threshold of $0.5$. On the test set, the two models were well-calibrated
%and had virtually identical predictive performance (AUC of $0.71$ and accuracy of $66\%$). This performance is comparable, if not superior, to the performance of many models
%deployed in real-world criminal justice settings 
%\cite{desmarais2020predictive}. 

\begin{figure}[t]
    \centering
    \includegraphics[width=1\linewidth]{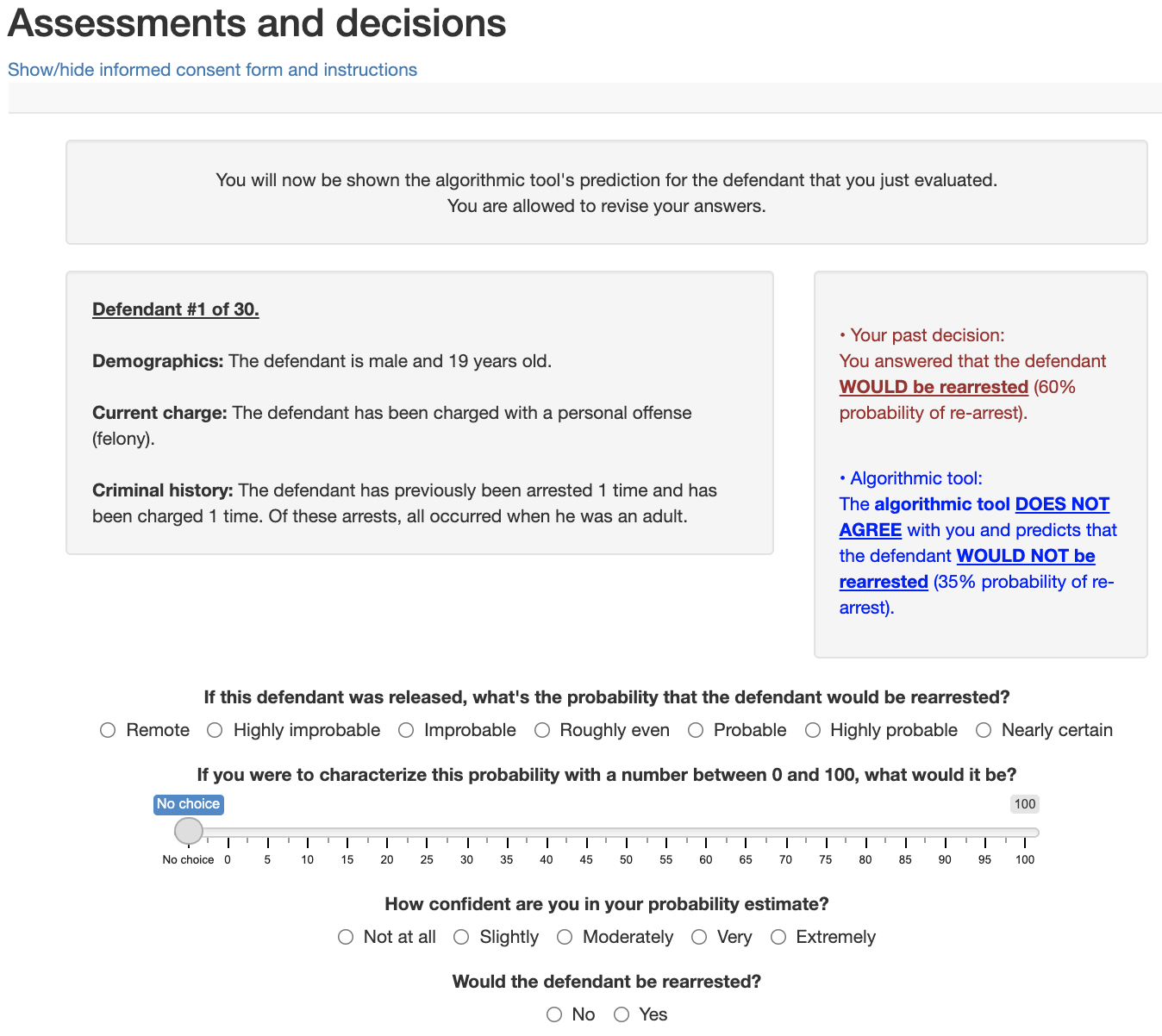}
    \caption{Sample vignette in the study. The panel on the right, which
    contained the AI's recommendation, became visible only once participants had
    made an initial prediction.}\label{fig:sample-vignette}
\end{figure}

Given these two algorithms, %we selected cases where either the two models agreed or they both disagreed. % Moreover, we were particularly interested in using cases of agreement (disagreement) that typically produced high (low) confidence judgments from humans. 
%We identified cases in the dataset released by \citeA{fogliato2021impact} where either (a) more than 80\%
%of the participants made the same prediction, or (b) less than 60\% of the participants made the same prediction. We then
%Specifically, 
we selected the instances of model-agreement as cases where the models both predicted the likelihood of re-arrest to be either above 70\% or below 30\% (48 cases), and the instances of model-disagreement as those where their binary predictions differed (34 cases). 
This sample of ``easy'' (high confidence, model agreement) and ``hard'' (low confidence, model disagreement) cases was then used to construct  the vignettes. %for participants.

Each vignette (see Figure \ref{fig:sample-vignette}) contained a description of the defendant with exactly
the same set of variables that were used for model training. Vignettes also possibly (see Experimental design below) included 
the output from one of the two algorithms---the estimated probability of re-arrest (e.g., 78\%) and binary prediction 
(re-arrest vs. no re-arrest). 

Given a vignette, participants were asked to estimate the likelihood that the defendant would be rearrested in the three years following release. Participants responded using a 7-point Likert scale (``Remote`` -- ``Nearly
certain''), and also a $[0,100]$ slider of probabilities (in increments of 5\%). They had to provide their confidence in the likelihood judgment (5-point
Likert scale ``Not at all'' -- ``Extremely'') and,  
finally, 
% since probability judgments made by participants may not map
% uniformly onto binary predictions \shortcite{fogliato2021impact}, 
they were also asked to predict whether the defendant would be rearrested or not (``no'' vs. ``yes'').

\subsubsection{Experimental design and procedures}
Tests of our three hypotheses require independent manipulation of the level of agreement between the participant and algorithm (to test for homophily effects); incentive structure (to test for cost effects); and availability of feedback (to control for learning effects). In order to avoid cross-condition learning, we used a fully between-participants design with  $18 = 3 \text{ \{Homophily\} } \times 3\text{ \{Incentive\} } \times 2\text{ \{Feedback\}}$ conditions.

After consenting, all participants were provided with instructions that described the task, showed an example case, and explained that the algorithm's
recommendations  had been generated by a model
that was 
% accurate and 
well-calibrated, including an explanation of what ``calibration'' meant.

The first phase consisted of $15$ cases, randomly drawn from the previously-described subset, with $10$ hard cases and $5$ easy cases. The re-arrest rate for these cases was matched to the re-arrest rate in the dataset (around $40\%$).
For each case, participants were first
asked to answer the four 
rating and 
prediction questions. After answering, the algorithm's recommendation was shown and participants were allowed to revise their answers. Two
attention checks were randomly inserted in this sequence. 
%Cases were shuffled in random order for each participant.
At the end of the first phase, participants were asked to provide their confidence in their predictive ability (5-point Likert scale from ``Very low'' -- ``Very high''), and whether they agreed that the model would help them make predictions (7-point Likert
    scale from ``Strongly disagree'' -- ```Strongly agree''). 

The Homophily manipulation modified the share of first-phase cases in which the algorithm's binary predictions
matched the participant's binary predictions. This manipulation focused on the hard cases, as the participant should be most uncertain and the models also generated different predictions. 
% say how we got data for the participants
For those $10$ cases, we ensured that the algorithm matched the participant's binary prediction on 9 (High homophily), 5 (Medium), or only 1 (Low) cases. Since these cases were ones where the models disagreed, we simply switched between models depending on which made a different binary prediction from the participant (see also \citeA[Experiment 3]{lu2021human}).\footnote{This design enabled us to truthfully tell participants that all algorithm predictions came from a well-calibrated model.}
This manipulation ensured that the model's
accuracy, as inferred by the participant, was orthogonal to the level of agreement.

We also manipulated whether 
outcome feedback was available during the
first phase. In particular, after receiving the algorithm's recommendation and
potentially revising their predictions, participants in the Feedback condition were 
informed of whether the defendant was actually rearrested, while those in the No feedback condition were not told anything about the eventual outcome. No participant received any feedback in the second phase of the survey. 

For the second phase of the experiment, we drew another random sample without replacement of $15$ cases (with $40\%$ re-arrest rate), this time with $8$ hard and $7$ easy cases. Only the predictions of the Lasso model were shown in this phase. The Incentive manipulation determined participant  compensation based on their performance in this phase. 
We used three different structures, all of which provided no reward for an incorrect (binary) prediction:
\begin{itemize}
    \item High true positive (High TP): $\$0.36$ for a true positive prediction (i.e., re-arrest) and $\$0.18$ for true negatives.
    \item Neutral: $\$0.27$ for each correct prediction. 
    \item High true negative (high TN): \$0.36 for true negative predictions (i.e., no re-arrest) and \$0.18 for true positives. 
\end{itemize}
Note that this incentive manipulation should only impact binary predictions, as those determine the payoffs. Participants  were presented with a detailed description of the incentives, including comprehension questions before and after the second phase. The incentive structure was also listed in each vignette as a reminder.

At the end of the second phase, participants again rated   confidence in their predictive ability, and whether they agreed that the model helped them make their predictions. A subset of participants\footnote{A bug in the survey meant that only a (random) subset of participants were asked this question.} were also asked whether they thought that their predictions had been
influenced by the incentives.

For data analysis purposes, we used participants' judgments and ratings, and also three additional measures adopted
from prior work \cite{yin2019understanding, green2019principles}:
\begin{itemize}
    \item \emph{Agreement fraction}: the fraction of cases in which the
    participant's binary prediction matched the algorithm's.
    \item \emph{Switch fraction}: out of the cases for which participant and algorithm initially disagreed, the fraction of cases where the participant changed binary prediction.
    \item \emph{Influence}: the median (per participant) difference between the revised and initial
    numerical likelihood estimates made by the participant, divided by the
    difference between the model and the participant's initial likelihood
    estimate. 
\end{itemize}

Our statistical analysis employs the average of each metric computed at the participant level.

\subsubsection{Participants}
A sample of $862$ participants was recruited on MTurk, all with HIT approval rating $>$90\%, $>$500 completed HITs, and physically present in the US. Participants were paid variable amounts depending on performance; mean compensation was \$4.40 (sd=\$0.60), translating to average payment of slightly more than \$10 per hour. $369$ participants failed the attention checks,\footnote{Because of the complexity of the task, we wanted to ensure that participants were actually paying attention. We thus used more severe attention checks (i.e., not simply ``click here to continue''). As a result, we had higher-than-normal failure rates.} resulting in a final sample of $N=493$.

% We also did not measure or test for order effects within each phase of the experiment. -> cases were shuffled in random order so there was no order effect

%% file: sections/results.tex
% \textbf{Note from DD: We need statistical tests (t-tests, binomial tests, etc.) for most of the claims in this paragraph.} 
The two measures of likelihood judgments---Likert scale and probability slider---were highly correlated ($\rho=0.77$, $p<0.001$), so we analyze only the probability judgments. 
We first consider the targeting effectiveness of our homophily manipulation. In particular, we assess the impact of this manipulation on participants'
judgements in the first phase of the survey.
We should expect to observe higher agreement between the
participants and the model's predictions on easy cases. 
Consistently, the average agreement fractions for
participants' initial and revised binary predictions on these easy cases were both above 80\%.
319 participants (64\%) agreed with the algorithm on all 5 easy cases, and 53 (11\%) agreed for 4-of-5
cases.
%, with higher agreement on high-confidence positive instances (i.e., re-arrests) than on negative ones (92\% vs 73\% for revised predictions). 
% Interestingly, participants
% tended to overpredict the occurrence of re-arrests, 
% and agreement on high-confidence positive instances (i.e., re-arrests) was higher than on negative ones (92\% vs. 73\% for revised predictions). 
Participants' predictions were more accurate for easy cases than difficult ones (classification accuracies were 73\% vs. 45\% respectively). Participants also reported being more certain about their
predictions on the easy cases (3.9 (out of 5) vs. 3.5; p-value of paired t-test$<$0.01). 
% In parallel
% with agreement, participants' confidence was slightly higher on positive cases compared to negative ones (4.2 vs. 3.6; p-value of paired t-test$<$0.01).
%Unsurprisingly, when the participant observed an AI's recommendation that
%matched their initial binary prediction, confidence in the prediction increased.
Lastly, participants spent longer on difficult cases (mean = 45s) than on easy cases (mean = 40s) for initial predictions (paired t-test, $p<0.01$).

\begin{figure}
    \centering
    \includegraphics[width=\linewidth]{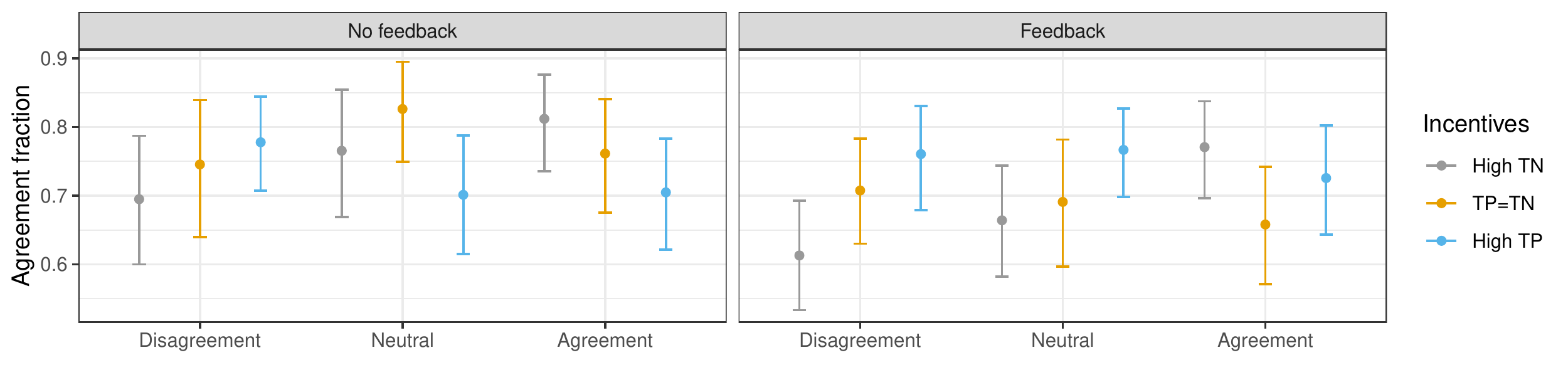}
    \includegraphics[width=\linewidth]{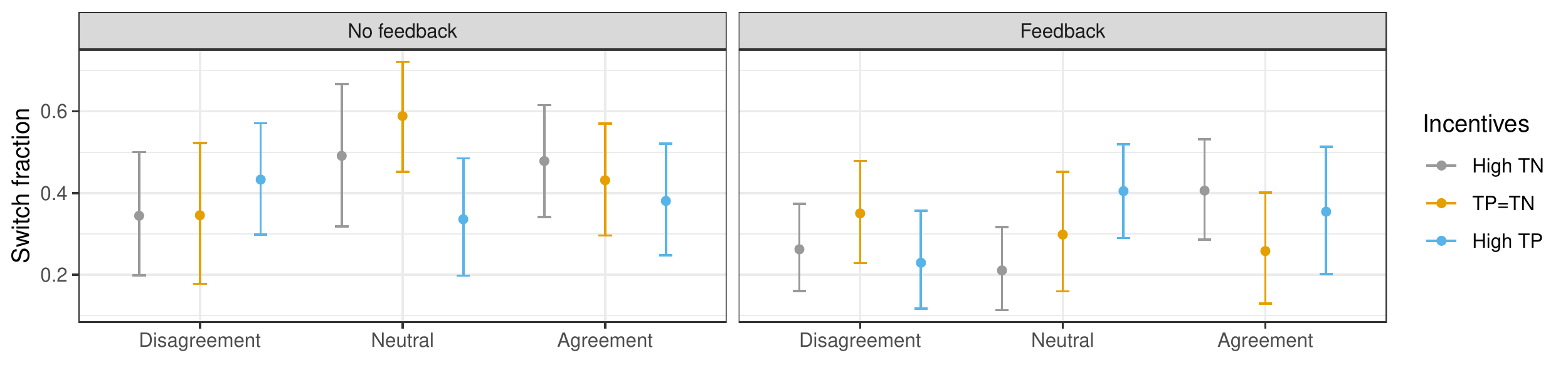}
  \includegraphics[width=\linewidth]{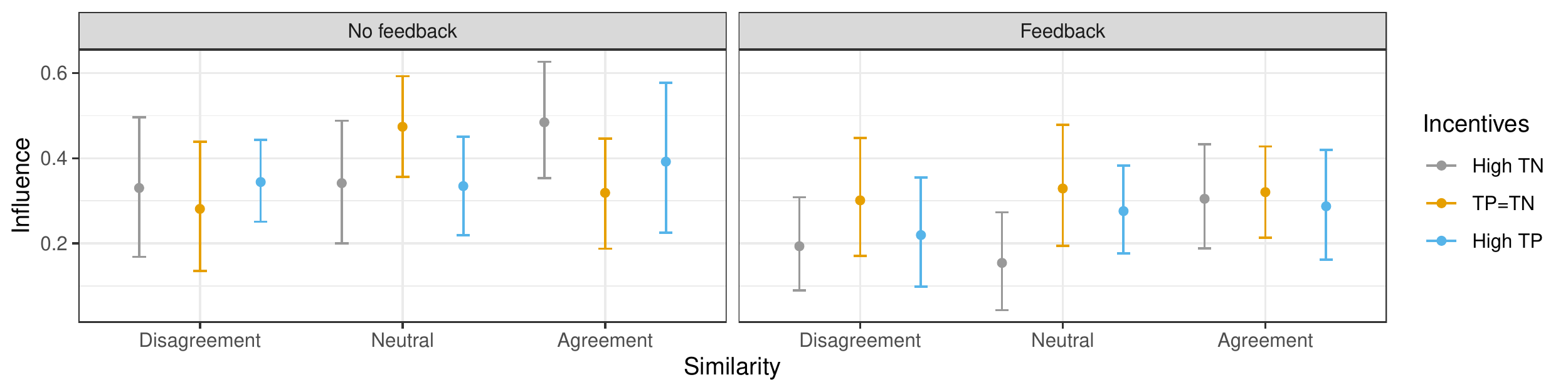}
    \caption{Second-phase agreement fraction (top), switch fraction (middle), and mean influence (bottom) for agreement (horizontal axis), incentives (color), and feedback (panels) manipulations (means and 90\% confidence intervals computed via nonparametric bootstrap and percentile method).}
    \label{fig:outcome_metrics_reliance}
\end{figure}

The cleanest tests of our hypotheses are based on responses in the second phase of the experiment, as the initial manipulations should have had an impact, and there should not be further learning since participants receive no feedback. 
Figure \ref{fig:outcome_metrics_reliance} shows the agreement fraction, switch fraction, and influence for participants' predictions across treatments in the second phase. 
We fitted two separate linear regressions---one to predict agreement fraction and one to predict switch fraction---with \texttt{homophily} and \texttt{incentives} as independent factors (the latter dichotomized into symmetric vs. asymmetric incentives to reflect H2b), interacted with the two \texttt{feedback} interventions.
We use robust standard errors of the coefficients estimates for hypothesis testing.

For the agreement fraction regression, the coefficients relative to homophily and incentives were all virtually zero and not
statistically significant. However, the regression results indicate that, controlling for incentives effects, the presence of outcome feedback drastically reduced the agreement fraction across all conditions, e.g., by 0.1 (0.76 vs. 0.66) in the low-homophily manipulation (Wald test of difference:  $p<0.04$). 
This pattern is also visible in Figure \ref{fig:outcome_metrics_reliance}. 
In the switch fraction regression, 
we find that the medium-homophily manipulation led to higher reliance on the model compared to the low-homophily one, but only when outcome feedback was absent (increase of 0.09 with one-sided $p<0.07$). 
While the presence of feedback substantially decreased reliance also according to this metric, 
the homophily interventions did not appear to impact reliance, e.g., the effect of medium homophily (vs. low) was even negative. 
We also find that the coefficients for the incentives effects were close to zero and not statistically significant. 
A parallel analysis for influence yields similar results: Neither higher levels of homophily nor asymmetric incentives increased reliance. However, these regression results again indicate that the presence of outcome feedback decreased reliance.

%We now turn to a brief exploratory analysis of interaction effects between agreement
%and incentives. In Section \ref{sec:experiment}, we hypothesized that an
%asymmetric incentives structure could make participants rely more heavily on the
%AI recommendations when the level of agreement in the initial phase of the survey was high. The patterns in
%Figure \ref{fig:outcome_metrics_reliance} suggest that this may not be the case:
%Participant's reliance does not seem to depend on the exact type of errors on which
%incentives are placed. For example, we observe that, for the participants in the
%high agreement condition, the agreement fraction for those with symmetric
%incentives is larger than that for those with high incentives on true negatives,
%but lower than for those in the remaining condition. However, given that
%participants tended to make positive predictions in the data (compared to the base
%rate), we would expect the opposite, i.e., reliance to be higher when they are
%incentivized to make negative predictions; we describe this
%pattern more in depth in Section \ref{sec:incentives_vs_predictions}. In addition, in presence of feedback the patterns that we observe
%are not entirely consistent across the three metrics.

Small effects combined with the limited sample size of our study may explain the null results discussed in the previous paragraph. 
As a confirmatory analysis,
we investigated whether any sizable effects could be detected on difficult cases, i.e., those where predictive uncertainty was high and participants were not confident about their predictions. Any effects should be most
salient on these subsets of cases, so we performed the same regression analyses for the three metrics. 
The conclusions from these analyses are analogous to those we have discussed above: Reliance increased from low to medium homophily only when outcome feedback was absent (here effects were statistically significant across all three metrics), and providing feedback decreased reliance. No other effects were statistically significant. 
%\footnote{To further validate our results, we conducted a post-hoc analysis using a mixed-effects model. This choice is motivated by the fact that previous studies have found substantive individual differences in people’s tendency to rely on social informational sources \cite{molleman2014consistent, toelch2014individual}. Accordingly, we fitted models (logistic for binary outcomes, linear otherwise) with random effects (intercepts) for the different workers.
%to account for their propensity to rely on the AI recommendation's. 
%Effects of the interventions were captured through fixed effects. We repeated the analysis using the a binary metric of agreement between the AI and the participant, for all cases and only for the cases on which the initial predictions didn't match separately, and the influence as outcomes, measured for each defendant's case. This analysis yielded similar results to those that we report in the paper.}
% TODO: actually do this analysis

Participants were also asked about perceived
utility and trust of the AI model in two questionnaires, at the end of the first phase and of the experiment respectively.
Results from these ratings somewhat mirror the findings
on the objective measures of reliance. 
We could only detect one significant effect for the medium-homophily intervention which increased perceived utility of the model over the low-homophily one in the first questionnaire in absence of outcome feedback (one-sided $p<0.04$). However, this effect (or any other) could not be detected in the questionnaire at the end of the survey. 

Prior studies have found that differential payment for types of performance can influence people's behaviors \cite{evans2004if}. No such effects were found in our experiment. In particular, participants
who received larger rewards for true negatives did \emph{not} make more negative predictions. 
We regressed proportions of
of positive predictions (i.e., of re-arrest) made by each participant before seeing the algorithmic recommendations on incentives manipulations. 
These shares were virtually identical across the interventions (all between 57\% and 59\%) and the differences based on Wald tests were not statistically significant. 
Additional analyses focused solely on low-confidence and high-uncertainty cases had similar results. In the final questions, 90\% of
participants reported the correct rewards for the two types of successes. Moreover, the conclusions of this analysis do not change even if we exclude the 10\% of participants who misreported the rewards structure.  60\% of the participants that encountered asymmetric rewards reported that their predictions were either slightly or not impacted at all by the incentives, and another 20\% reported that impact had been moderate.

Finally, we analyze participants' confidence in the predictions using the regression analysis described earlier. 
In the first questionnaire, the medium-homophily intervention positively impacted participants' confidence ratings compared to the low-homophily one when feedback was absent (effect is 0.38 on a 1--5 Likert scale rating; Wald test $p<0.01$). The high-homophily manipulation also increased confidence compared to the medium-homophily one, but the effect was smaller (0.17, one-sided $p<0.08$).
In presence of outcome feedback, no effect was detected. 
Similar results were found for the same question inserted in the final questionnaire, although the effects relative to the homophily interventions were slightly smaller. 
A regression analysis of the confidence ratings reported by participants for each of the predictions (here the average rating by participant) in the second phase delivered similar results: Higher levels of homophily increased confidence in absence of feedback (both one-sided $p<0.07$), but this was the only effect that could be detected.

\subsection{Remarks}
Almost none of our initial hypotheses were confirmed by this experiment. H1 and H3 imply that participants who experience greater homophily (H1) and no feedback (H3) should have the highest levels of reliance. At the same time, participants should rely
more on the recommendations when the cost of prediction errors is asymmetric (H2b). 
Although some of the results suggest the presence of these effects, there is no clear impact of the sort predicted by the hypotheses. 
The non-monotonicity of the impact of homophily is particularly surprising, as in multiple instances the largest effects were found for medium levels. One possibility is that the high and low levels led participants to largely disregard the AI recommendations, though for different reasons: The former found the AI to be redundant, while the latter found the AI to be error-prone. 
%We similarly do not find any strong effects of incentives on participants' patterns of judgments (H2). Moreover, we find largely-null findings even though we have a reasonable sample size (roughly 27 participants per condition, each providing multiple judgments). %given previously reported effect sizes for experiments such as this one.

%% file: sections/discussion.tex
Our results have provided little or no support
for the key hypotheses driving our research study: 
Although homophily and incentives 
impact human reliance on other people, 
they do not (in this setting) 
seem to strongly influence 
human reliance on AI tools. 
Participants who were shown AI recommendations 
that matched their predictions more often
did not appear to rely or trust the tool 
substantially more than their counterparts (H1).
Agreement between the AI tool and participants
did translate into higher subjective confidence for the participants,
but not higher usage of the AI information. 
Incentives for different types of predictions 
did not affect participants' judgments 
or their reliance on the recommendations (H2). 
However, the presence of outcome feedback 
did decrease participants' reliance on the tool (H3). 

Homophilic effects %---individuals tend to associate with similar others---
are prevalent in social life, influencing
individual associations~\cite{golub2012homophily} and trust relations~\cite{tang2013exploiting} in social networks.
The absence here of effects of homophily---operationalized 
as similar decisions over prior cases---could 
be explained in at least two interrelated ways. 
One possible explanation is that some factors that
impact inter-human trust relations 
simply do not transfer to the human-AI case~\cite{glikson2020human}. 
From this perspective, homophily 
is relevant to understanding 
how agents are influenced by other humans, 
but not relevant to the algorithmic decision support case. 
\citeA{mahmoodi2018reciprocity}, for example, have highlighted the
significance of \textit{reciprocity} 
for understanding social informational influences, 
finding agents to be more open to influence
from partners who are expected to \textit{reciprocate} this influence later on.
Importantly, \citeA{mahmoodi2018reciprocity} also find 
that this dynamic process of reciprocity was 
``abolished when people believed that they interacted with a computer,'' 
potentially because people expected 
that algorithms would not (or cannot) reciprocate. 

A second, not exclusive, explanation derives from the
many dimensions of ``similarity'' among individuals, 
including demographics such as race and gender,
underlying values, political affiliations,
shared attitudes and beliefs, and more. 
Most of these factors can potentially 
drive the emergence of homophilic effects \cite{monge2003theories}, 
but not all of them will be relevant 
in a specific context \cite{ahmad2011trust}.
This second, narrower explanation allows for the possibility 
that people could experience homophilic effects with an algorithm, 
but only if there were appropriate perceived similarities with the algorithm.
For example, this explanation would leave it open 
that people could experience consistent homophilic effects 
if they knew that the algorithm had been developed 
by someone who shared their values. 
Both of these potential explanations 
provide avenues for future research.
In either case, though, our findings provide reasons
to be cautious when transporting findings 
about interpersonal relations 
from psychological and organizations sciences 
to the case of human-AI interaction.
% homophily observed in many fields

% but there are many different dimensions (identity, values, beliefs, ...) and
% not all these dimensions are always present and depend on the mode of
% interaction (e.g., real life vs. gaming platform). So in the case of
% human-algorithm interaction, our work provides reason to think that agreement
% on cases is not sufficient for eliciting feelings of homophily

% [Riccardo] low quality data collected on mturk because of the platform's
% market structure. Cite studies of bad experiences of MTurk workers. Recall (or
% present) the results on time spent on each assessment by participants in our
% survey. Argue that future studies should understand how data of better quality
% can be collected. too little money

% [Riccardo] even if the data were of reasonable quality and we had found
% stronger effects, would our results generalize to real world decision-making?

One might also worry about the pool of participants. 
Researchers have long cautioned against the potentially low
quality of data that are collected through crowdsourcing experiments such
as ours, particularly those on MTurk 
\cite{paolacci2010running, kennedy2020shape}.
The platform itself incentivizes requesters 
(i.e., people conducting experiments)
to offer low payments and workers (i.e., participants) 
to exert minimal effort. 
Thus, workers often adopt a variety of strategies 
to maximize profit \cite{mcinnis2016taking, chandler2014nonnaivete}, 
such as doing multiple HITs simultaneously. 
This worker strategy would not necessarily be an issue
if we were conducting, for example, a quick five-question survey. 
The present experiment is complicated, however, 
and requires people to draw relatively fine distinctions. 
We attempted to minimize these risks by requiring 
a higher worker approval rating 
than for many other studies that are similar to ours
\cite{green2019principles, green2020algorithmic}. 
We also  employed rigorous attention checks 
(and had correspondingly higher failures of those checks). 
We thus expect that we likely filtered out
a large share of low-quality responses, 
and so that possibility is less likely to explain our null results.

Another alternative hypothesis is that our study participants 
could have performed numerous tasks similar to ours in the past,
or have strong prior expectations about the possibility 
that an AI could be helpful for these kinds of decisions.
Their (already mature) beliefs about AI tools 
may have not been influenced by the short interaction 
that they had with our tool. 
However, the positive results around the impact of homophily on
\emph{confidence} do not appear to be compatible with these two possibilities. 
Similarly, the lack of an effect of incentives on predictions 
could be due to the fact that
the offered rewards were too small to nudge participants to change their
predictions. Indeed, previous studies have reported that payments do not
significantly increase the quality of the data that are collected
\cite{buhrmester2016amazon}. In our experiment, despite information about the
base rate and feedback (for some participants), participants may not have 
realized that their rewards would have been likely higher had they given the same
answers in all assessments. 
Interestingly, however, we could not detect any effect of this manipulation, even
on the the cases on which participants were less confident about their
predictions. 

%Lastly, we draw a connection 
%Even once we account for the fact that a large fraction of the data that we collected are in \citeA{poursabzi2021manipulating},

% Our study findings raise serious doubts 
% the questions of whether our study findings would generalize to real-world decision-making remains unclear. We see (at least) two points of disconnect between the predictions made by laypeople in our study and expert decision-making. First, our participants predicted the occurrence of a re-arrest, whereas ... % add 

Future research should carefully take into account and address
the key limitations of our study. 
In particular, our experimental results have shown 
that small monetary bonuses tied to accuracy
do not promote changes in crowdworkers' behavior. 
We actually found the same result in a pilot study on predictions of loan repayment, and so increased the incentives in this experiment to see whether any effect would be revealed.
If our null finding were replicated in other experimental setups, 
then we would have further evidence that incentives may not represent valid proxies 
for context-dependent costs in real-world decision-making. 
Consistent with the findings of \citeA{lu2021human}, 
our experiment has also highlighted that crowdworkers' trust and reliance on AI tools may be insensitive 
to interventions on their level of agreement with the recommendations generated. 
Alternative study designs may achieve more promising results, for instance by increasing the duration of the interaction of AI and participant while keeping them fully engaged in the task. Lastly, the connection between confidence and homophily uncovered in our experiment represents another interesting research direction. 

% Even if we had found sizeable effects 
%of our interventions on participants' behavior, 
%there still remains the question

% Why do our results (partially) diverge from those of \citeA{lu2021human}?
% As we have mentioned in \textsection\ref{sec:background}, the two experimental designs differ.
% Second, in our experiment participants were provided with both the binary and probability predictions of the model. 

% Similarly to our results, \citeA{poursabzi2021manipulating} found that 
% even though some of the participants may report being more confident in the model's predictions, they may not rely on its recommendations more heavily.  

% TODO: argue that the fact that we also show predicted probability might have less of an effect

% limitations
% our interventions were not effective

% 